# Photon: New light on an old name

Helge Kragh[*]

**Abstract**: After G. N. Lewis (1875-1946) proposed the term "photon" in 1926, many physicists adopted it as a more apt name for Einstein's light quantum. However, Lewis' photon was a concept of a very different kind, something few physicists knew or cared about. It turns out that Lewis' name was not quite the neologism that it has usually been assumed to be. The same name was proposed or used earlier, apparently independently, by at least four scientists. Three of the four early proposals were related to physiology or visual perception, and only one to quantum physics. Priority belongs to the American physicist and psychologist L. T. Troland (1889-1932), who coined the word in 1916, and five years later it was independently introduced by the Irish physicist J. Joly (1857-1933). Then in 1925 a French physiologist, René Wurmser (1890-1993), wrote about the photon, and in July 1926 his compatriot, the physicist F. Wolfers (ca. 1890-1971), did the same in the context of optical physics. None of the four pre-Lewis versions of "photon" was well known and they were soon forgotten.

## 1. Introduction

Ever since the late 1920s, to physicists the term "photon" has just been an apt synonym for the light quantum that Einstein introduced in 1905. Although "light quantum" is still in use, today it is far more common to speak and write about "photon." The name is derived from Greek (φώτο = photo = light) and the "-on" at the end of the word indicates that the photon is an elementary particle belonging to the same class as the proton, the electron and the neutron. The popularity of the name is illustrated by a Google search for either "photon" or "photons," which results in about 14.2 million hits, reduced to 3.5 million hits on Google Scholar. Not only is the term universally adopted by physicists, chemists and engineers, it has also found its way into commercial and cultural contexts. There are even poems devoted this most common particle in the universe. Here is one:[1]

[*] Centre for Science Studies, Department of Physics and Astronomy, Aarhus University, Aarhus, Denmark. E-mail: helge.kragh@ivs.au.dk. Parts of this note is included in a larger manuscript, dealing also with the origin of the names "plasma" and "fission," which has been submitted to *European Physical Journal H*.

> A photon he could only surmise, has broken through his glass.
> Her light has scattered through his gauzy prism.
> It has filled his void with notions of the intimate, the unknown, the truth.
> Brimming into the depths of his silica, she disperses.
> And yet she comes together, concentrated, at one focal point.
> Concave he is by nature, he allows her to assimilate at the core of his essence.
> For once she has ignited him, they shine.

This paper is not concerned with the meaning and properties of the photon, but solely with the origin of its name and the immediate response to it in the period from about 1927 to 1932. In almost all accounts it is stated that the name was originally coined by the American chemist and physicist G. N. Lewis in 1926, although in a sense that differs from the one associated with the light quantum. There is a good deal of truth in it, but it is far from the whole truth. It is hard to know when, exactly, the word "photon" was first used in a scientific context, but as far as I can tell it was in 1916, ten years before Lewis reintroduced it. This earlier history had little or no impact on the later development, so it is perhaps understandable that it has not been much noticed, if at all. Yet the early appearance of "photon" in the context of visual science and physiology is interesting in its own right and deserves to be known by historians of physics.

**2. Einstein's light quantum**

The origin and early development of the idea of the photon, in the sense of a localized quantum of electromagnetic radiation, is well known and described in an extensive literature [e.g., Kidd, Ardini and Anton 1989; Pais 1982, pp. 372-388, 402-414; Stuewer 2006]. To summarize, in his classic paper in *Annalen der Physik* of 1905 Einstein proposed that free monochromatic radiation of frequency $v$ was composed of "energy quanta" given by $E = hv$, an expression he only wrote down in this form the following year. Although he was in this way able to explain in a simple way the photoelectric effect and Stokes' rule of photoluminescence, the *Lichtquantenhypothese* was generally resisted by the large majority of physicists.

      The response to the light or energy quantum did not change significantly after Einstein in 1917 developed his theory by assigning a momentum of $p = hv/c$ to

---

[1] https://medium.com/poetry-prose/a841fb7935cb

the quantum. Only then did it have the properties of a real particle, which is the reason why Abraham Pais [1982, p. 408] speaks of Einstein's photon as dating from 1917. Of course Einstein did not use the name "photon" at the time, and nor did he use it in his later publications.

The main reason for the acceptance of the light quantum during the years 1922-1925 was the famous series of experiments on X-ray scattering that Arthur H. Compton conducted in late 1922 and for which he was awarded the Nobel Prize for 1927 [Stuewer 1975a]. Incidentally, in his two papers of 1923 reporting the results of what soon came to be known as Compton scattering, one in *Physical Review* and the other in *Philosophical Magazine*, he did not refer to Einstein's light quanta and neither did he mention Einstein by name. Nonetheless, the experiments did vindicate Einstein's theory and were, in this respect, a point of no return. By 1925-1926 the radiation quanta were generally accepted and conceived as elementary particles, no less real than the electron and the proton. With the emergence of the quantum theory of radiation, developed soon thereafter by Paul Dirac and Pascual Jordan, the light quanta were shown to fit into the general theory of quantum mechanics. Now indispensable, they were increasingly called photons.

## 3. Successful name, failed concept

The source of the name "photon" that entered physicists' vocabulary is the late 1920s was Gilbert Newton Lewis, professor at the University of California, Berkeley. However, while Lewis' name of 1926 was widely adopted, his concept behind the name was totally ignored and is today known only by a few historians of physics. Lewis was a distinguished physical chemist particularly known for his innovative work in chemical thermodynamics and the structure of atoms and molecules [Lewis 1998; Coffey 2008, pp. 113-149]. His theory of a shared pair of static electrons making up the covalent bond in molecules is widely seen as an anticipation of the later valence model based on quantum chemistry [Stranges 1982].

Apart from being a brilliant chemist Lewis also had a deep interest in theoretical physics, although in this area he generally preferred to follow his own, somewhat heterodox ideas rather than those of mainstream physics. In 1925 he became interested in the conceptual problems of radiation theory which he attempted to solve by postulating as a fundamental principle that time is symmetric. According to Lewis' picture of the radiation process, emission and absorption of light occurred in complete symmetry. "An atom never emits light except to another atom," he wrote. "It is as absurd to think of light emitted by one atom regardless of the existence of a receiving atom as it would be to think of an atom absorbing light without the existence of light being absorbed" [Lewis 1926a, p. 24; Stuewer 1975b]. The emission process would be the exact reverse of the absorption process, for which reason the two separate processes could be condensed into a single one, transmission. It was in this context that he proposed in a paper in *Nature* dated 29 October 1926 that instead of the light quantum one should consider "a new kind of atom" or what he called a "photon" as the carrier of light [Lewis 1926b].

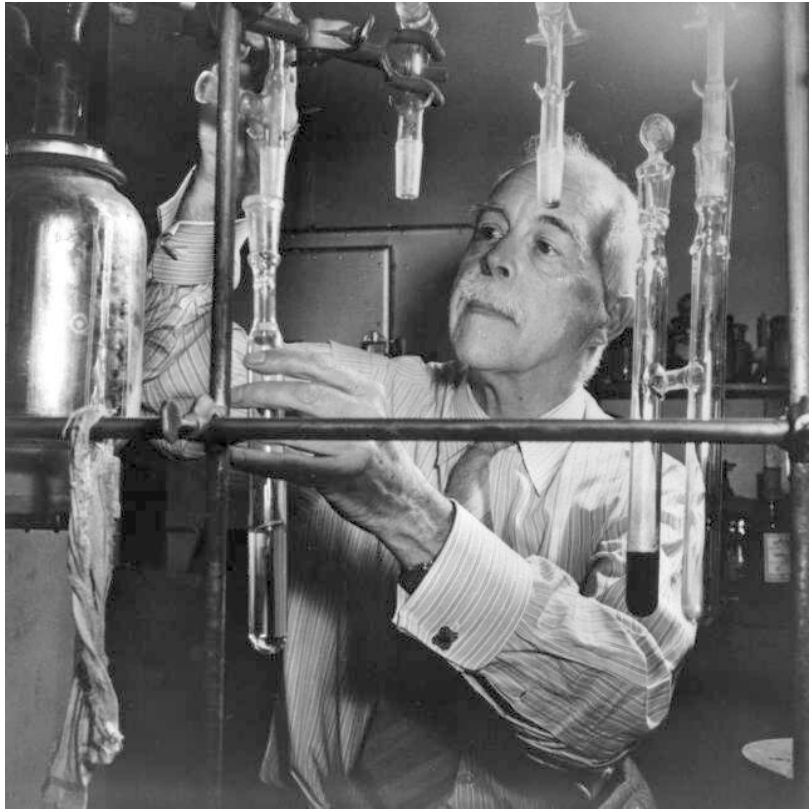

Fig. 1. Gilbert N. Lewis in his laboratory, 1944.

Contrary to Einstein's light quantum, Lewis' photon was a conserved quantity in the sense that in an isolated system the total number of photons would remain constant. The photon was uncreatable and indestructible, unlike the modern photon but much like John Dalton's immutable atoms of the past: "We might as well attempt to introduce a new planet into the solar system, or to annihilate one already in existence, as to create or destroy a particle of hydrogen" [Dalton 1808, p. 212]. Replace "particle of hydrogen" with "photon," and you have Lewis' claim. Further arguing that "one, and only one, photon is lost in each elementary radiation process," Lewis [1926b, p. 875] wrote:

> It would seem inappropriate to speak of one of these hypothetical entities as a particle of light, a light quantum, or a light quant, if we are to assume that it spends only a minute fraction of its existence as a carrier of radiant energy, while the rest of the time it remains as an important structural element within the atom. It would also cause confusion to call it merely a quantum, for … it will be necessary to distinguish between the number of these entities present in the atom and the so-called quantum number. I therefore take the liberty of proposing for this hypothetical new atom, which is not light but plays an essential part in every process of radiation, the name *photon*.

Lewis' photon thus was very different from the corpuscle of radiant energy that Einstein had introduced. It is misleading to write about "Einstein's fundamental light-quantum, which Gilbert N. Lewis had christened the 'photon' in October 1926" [Mehra and Rechenberg 2000, p. 241].

Realizing that his photon was unconventional and hypothetical, in a couple of later papers Lewis sought to vindicate his hypothesis by showing that it, or rather the direct-interaction picture on which it was based, led to the correct expressions for blackbody radiation [Stuewer 1975b]. Nobody listened to him, and within a year or so he seems quietly to have abandoned the unorthodox idea. For a period of more than forty years his paper in *Nature* was only cited twice in scientific journals, in both cases in 1927 and by himself. The next time it was cited was in 1970, in a historical review article on the names of particles in physics [Walker and Slack 1970], although four years earlier also Max Jammer referred to it in his pioneering monograph on the history of quantum mechanics [Jammer 1966, p. 30].

In 1927, William Band, a physicist at the University of Liverpool, referred critically to Lewis' theory of "the emission and absorption of photons." However, his reference was to an earlier paper that did not include the term and not to Lewis' paper introducing "photon" [Band 1927].

**4. French connections**

More than two years before Lewis' article appeared in *Nature*, the French biochemist and physiologist René Wurmser made use of the name "photon." He was at the time well acquainted with the theory of photochemical reactions from the point of view of the light quantum hypothesis. This was a topic that Einstein had investigated in a paper of 1912, and Wurmser, collaborating with Victor Henri, verified at an early date Einstein's prediction that only a single light quantum of a given energy was needed to initiate a photochemical reaction [Henri and Wurmser 1913].

In September 1924 he submitted a paper to the *Annales de Physiologie* in which he reported his investigations of photochemical reactions relating to the role of chlorophyll in photosynthesis. He called into question the prevailing view that the oxygen produced in photosynthesis was derived from light acting on carbon dioxide. To explain his results he referred to the idea of energy transfer by molecular resonance suggested by his compatriot, the physical chemist and Nobel laureate Jean Perrin [Berberan-Santos 2001]. According to Wurmser, one had to assume that "the activation of a molecule, in the sense of J. Perrin, demands the absorption of an integral but variable number of *photons*" [Wurmser 1925a, p. 60, emphasis added; Wurmser 1925b].

Apart from associating the photon with the energy $h\nu$, Wurmser did not comment on the name or use it more than once. It was just there, apparently unnoticed by his colleagues. Wurmser went on to a distinguished career in biophysics and the new science of molecular biology. At the age of 97 he referred to his old paper, noting that it contained several considerations of interest [Wurmser 1987]. The name "photon" was not among them.

Very little is known of the French physicist Frithiof Wolfers, who apparently changed his first name to Fred and signed his many papers as just "F. Wolfers." From about 1920 to 1940 he worked at the University of Algiers, after which he was appointed professor of physics in Paris. A prolific author of books and articles, his research focused on optics, thermoelectricity, piezoelectricity and X-rays. Working in part in Marie Curie's laboratory in Paris, in the 1920s he conducted an extensive series of experiments on the diffraction and diffusion of X-rays. In 1929 he published a book on nuclear physics and the transmutation of elements that carried a preface by Jean Perrin [Wolfers 1929].

When Compton reported his discovery of the effect that soon became named after him, Wolfers remained unconvinced of Compton's explanation in terms of light quanta (Friedel and Wolfers 1924; Stuewer 1975a, p. 261). He concluded from his own experiments that the observed phenomena amounted to an *experimentum crucis* in favour of the Bohr-Kramers-Slater (BKS) theory proposed by Niels Bohr and his collaborators Hendrik Kramers and John Slater in 1924. "My experiments completely confirm Bohr's thesis," he wrote [Wolfers 1925, p. 366; Wolfers 1924]. The physicists in Copenhagen seem to have been unaware of Wolfers' claim of having confirmed their hypothesis.

Given that the BKS theory was proposed as an alternative to Einstein's (and Compton's) theory of light quanta in the area of radiation-matter interaction, it is a little surprising that Wolfers [1926], in a paper communicated to the French Academy of Science by Aimé Cotton on 26 July 1926, introduced the name "photon" for what he also called a light quantum or an "atom of light":

> I shall use the name *photons* for the projectiles that supposedly transport radiant energy and possess the character of a periodic frequency $v$ (atoms of light). I then suggest that the photons may be repelled by the atoms of matter when they pass them closely … One may imagine that the repulsion is due to a kind of resonance between the photons and the resonators …

Like Wurmser, but in a very different context, Wolfers thought that his hypothesis was related to Perrin's idea of molecular resonance or what he called *induction moléculaire*. He published several papers on his experiments and hypothesis over the years, but only returned to the name "photon" in 1928. His paper of 1926, in which he used the name several months before Lewis, went as unnoticed as Wurmser's. The Web of Science lists no citations to it. On the other hand, Wolfers' work may have been well known in the French physics community. None other than Louis de Broglie, in his famous dissertation of 1924 marking the beginning of wave mechanics, referred to it [De Broglie 1925, p. 99].

**5. An apt name**

Whereas Lewis' concept was promptly forgotten, his and Wolfers' name was not. "Photon" was quickly and unceremoniously assimilated as an alternative name for Einstein's light quantum, and by the mid-1930s it had become the preferred one. Figure 2 and the following count of titles of scientific papers including photon(s) and light quantum (quanta) gives an impression of the popularity of the name proposed by Wolfers and Lewis:

|  | *1926-1935* | *1936-1945* | *1946-1955* |
|---|---|---|---|
| Light quantum (quanta) | 20 | 0 | 5 |
| Photon(s) | 19 | 29 | 243 |

The numbers are based on the Web of Science, cp. the caption to Figure 2.

As early as 1928, in the proceedings of the famous fifth Solvay Conference of October 1927, the term "photon" appeared in the title of the prestigious publication – *Électrons et Photons*. Preparations of the conference started in April 1926, before Wolfers and Lewis had suggested the term, and in the material leading up to the conference one looks in vain for any mention of "photon" [Bacciagaluppi and Valentini 2009, pp. 8-22]. Most likely the name entered the title of the proceedings only during the last phase of preparation, reflecting that several of the speakers and discussants used "photons" rather than "light quanta" in their reports. Among them were H. A. Lorentz, Louis de Broglie, Paul Dirac, Léon Brillouin, Paul Ehrenfest and Arthur Compton. It was only the latter who called attention to the origin of the new name [Compton 1928a, p. 57; 1928b, p. 156]:

> In referring to this unit of radiation I shall use the name "photon" suggested recently by G. N. Lewis. This word avoids any implication regarding the nature of the unit, as contained for example in the name "needle ray." As compared with the term "radiation quantum" or "light quantum," this name has the advantages of brevity and of avoiding any implied dependence upon the much more general quantum mechanics, or upon the quantum theory of atomic structure.

Although Lewis' photon was mentioned only by Compton, his time-symmetric theory of light was discussed by Brillouin, who argued that a certain photon paradox posed by Lewis would disappear with de Broglie's conception of "the trajectory of a photon" [Bacciagaluppi and Valentini 2009, pp. 369-370]. Brillouin and de Broglie were, most likely, aware of Wolfers' photon, but they did not refer to it. On the other hand, in the proceedings of the Solvay meeting the Belgian physicist Théophile de Donder referred to Wolfers' experiments to which Brillouin had drawn his attention [Bacciagaluppi and Valentini 2009, p. 466].

Compton promoted the new term at an early date, if sometimes in a way that suggested it was his own and not Lewis' invention (he may have been unaware of Wolfers' contribution). Thus, in his Nobel lecture of 12 December 1927 he referred to X-rays as "light corpuscles, quanta, or, as we may call them, photons" [Compton 1928c, p. 84]. Compton may also have been responsible for introducing the name in the popular science literature. In a paper in *Scientific American* from February 1929 he wrote about X-rays that they consisted of "photons," indicating the novelty of the name by placing it in citation marks. He rated the new particle of light highly [Compton 1929, p. 236]:

> The light which makes the plants grow and which gives us warmth has the double characteristics of waves and particles, and is found to exist ultimately of photons. Having carried the analysis of the universe as far as we are able, there thus remains the proton, the electron, and the photon – these three. And, one is tempted to add, the greatest of these is the photon, for it is the life of the atom.

9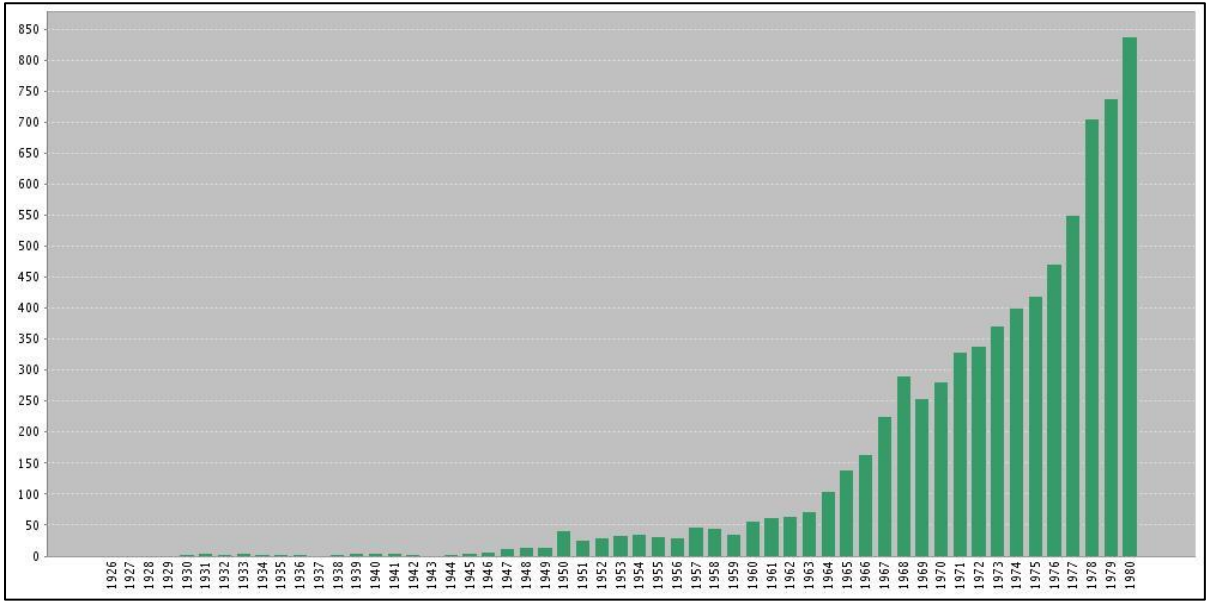

Fig. 2. The rise of the photon. Number of papers 1924-1980 in English with "photon" or "photons" in title. Total number = 7,325. Source: Web of Science. The scarcity of papers in the period 1936-1945 probably reflects the effect of World War II on academic physics publications. Users of Web of Science and similar databases should be aware that all titles appear in English translations (and indeed often mistranslations). A search for "photons" in the title thus results in a list with several papers before 1926, simply because the German *Lichtquanten* (or the French *quanta de lumière*) appears in modernized translations.

"Photon" became disseminated to the young generation of physicists through the early textbooks in quantum mechanics. The first two textbooks, both published in 1928, were by the Cambridge physicist George Birtwistle and by the Austrian physicist Arthur Erich Haas, at the University of Vienna. While Birtwistle [1928] avoided the new name altogether and stuck to "light quanta," Haas used both. In the beginning of his book he referred to "the radiation atoms that are at present generally called *light quanta* or, most recently, *photons*" [Haas 1928, p.6]. Hermann Weyl's *The Theory of Groups and Quantum Mechanics* was a translation of the German original published in 1929. In the English edition Weyl [1931, p. 42] introduced early on Einstein's idea of "a light quantum or photon," thus equating the two concepts. In the rest of the book he kept to "photon."

Even more influential was Dirac's classic *Principles of Quantum Mechanics* which started right away with describing light as composed of "small particles, which are called photons" [Dirac 1930, pp. 1-2]. He kept consistently to the name, without mentioning "light quanta" or the recent origin of the name "photon." It is unknown whether Einstein, the father of the light quantum, liked the name or not, but he never used it.

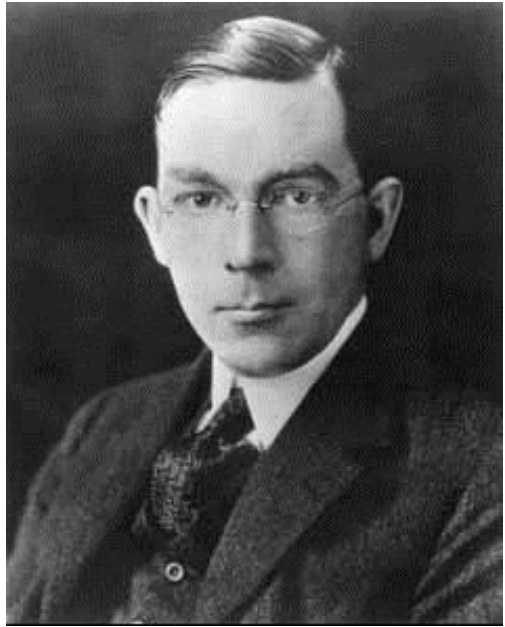

Fig. 3. Leonard Troland.

**6. The unnoticed genesis of "photon"**

Probably unknown to Lewis and almost all contemporary physicists, the word "photon" can be found in the scientific literature as early as 1916. It was coined by the American physicist and psychologist Leonard Thompson Troland, who used it as a unit for the illumination of the retina. Although little known today, and if known at all then for his work in experimental psychology, at the time he was considered one of America's most promising scientists. When he died tragically and prematurely in 1932 by a fall from the summit of Mount Wilson in California, his death was mourned in obituaries in *Science* (vol. 76, pp. 26-27) and *American Journal of Psychology* (vol. 44, 817-820).

A versatile and respected scientist, Troland followed parallel careers in psychology, physics and engineering. With Daniel Comstock, a physicist at Massachusetts Institute of Technology, he published in 1917 a semi-popular book on the modern theory of atoms, electricity and radiation. A section on the quantum theory of radiant energy included a discussion of "light atoms" or "the modern doctrine of light 'quanta'," of which Troland, who was the author of the section, clearly was in sympathy [Comstock and Troland 1917, pp. 182-189]. In 1922-1923 he served as president of the Optical Society of America. In addition, he had some training in biochemistry, and in 1916 he proposed one of the earliest theories of the chemical origin of life on Earth [Troland 1916a].

Troland studied psychology at Harvard University from where he obtained his doctorate in 1915 with a thesis on visual adaption, a line of work he continued over the years. His main work was the massive *Principles of Psychophysiology* that appeared in three volumes 1929-1932. He was particularly interested in photometric measurements of light impinging on the human eye, and it was in this context that he suggested the term "photon."

In an article dated 29 March 1916 on the intensity of light stimulating the eye, Troland introduced "photon" as a unit for physiological stimulus intensity, defining it as follows [Troland 1917, p. 32; Fig. 4]:

> A *photon* is that intensity of illumination upon the retina of the eye which accompanies the direct fixation, with adequate accommodation, of a stimulus of small area, the photometric brightness of which … is one candle per square meter, when the area of the externally effective pupil … is one square millimeter. The *physiological intensity of a visual stimulus* is its intensity expressed in photons. The photon is a unit of illumination, and hence has an absolute value in meter-candles. The numerical value of the photon, in meter candles, … will obviously be subject to some variation from individual to individual.

Troland first suggested the photon in a presentation given to the tenth annual meeting of the Illuminating Engineering Society in Philadelphia 18-20 September 1916. "I have," he said, "found it very convenient to express all intensity measures in terms of a unit retinal illumination which I have called the *photon*" [Troland 1916b, p. 950]. In the discussion following his talk, he mentioned as an advantage of the new unit that "the photon unit does not require so much mathematics, and I have been interested primarily in helping the psychologists, many of whom are studying vision somewhat at random."

In a later monograph on visual science published by the National Research Council, Troland followed up on his research and promoted the use of the new unit [Troland 1922]. However, although he and a few other authors continued to use "photon" for a period of time the unit never was widely used. By the mid-1920s it seems to have fallen into oblivion. On the other hand, it later gave rise to the eponymous unit "troland" (Td) which relates to the candela unit (cd) by

$$1 \text{ Td} = 1 \text{ cd/m}^2 \times 1 \text{ mm}^2 = 10^{-6} \text{ cd}$$

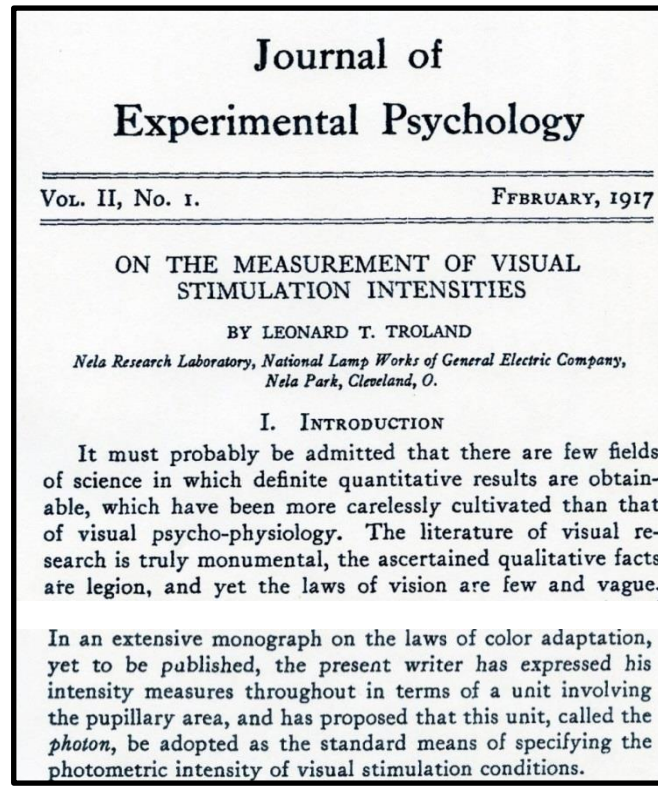

Fig. 4.

The Td unit is still used in physiological optics [Barten 1999, pp. 60-62]. Troland was aware of the discussion in physics concerning the light quantum hypothesis, but to him the photon was a quantity belonging to a different realm.

**7. Yet another origin of "photon"**

Troland was not the only scientist who suggested the name "photon" before Lewis. Curiously, also this second appearance of the name had its origin in the science of vision, in this case in an attempt to explain colour vision.

  The Irish physicist John Joly, professor of geology and mineralogy at Trinity College, Dublin, was – not unlike Troland – a remarkably versatile scientist [Nudds 1986]. He is probably best known for his estimate of 1899 that the age of the Earth was at least 100 million years, much longer than argued by Lord Kelvin on the basis of thermodynamical calculations [Burchfield 1990]. While his estimate of 1899 was based on geological evidence, when methods based on radioactivity became known he was quick to make use of them. In 1903 he drew attention to the importance of radioactivity as a source of terrestrial heat. Joly not only did important work in radioactivity (which included collaboration with Ernest Rutherford), he also developed a method for treating cancer by means of radium and generally had an abiding interest in medicine and physiology. About 1920 he turned his interest toward the theory of vision, and colour vision in particular. As a result, he developed what he called a "quantum theory of vision" with the aim of explaining the sensation of light as stimulated by the action of photo-electricity [Joly 1921a].

According to the view of Joly, light in the form of "light quanta" liberated and activated electrons in the visual fibres. Depending on their kinetic energy these photoelectrons would discharge one or more units of energy into the brain's cortex, thereby giving rise to a sensation of light. What he called a "photon" was a unit of light stimulus or sensation [Joly 1921b, p. 26 and p. 29]:

> The unit light stimulus discharged by a single visual fibre … must not be confused with the quantum which plays the part merely of the finger of the trigger. This minute quantity of energy discharged into the cerebral cortex evokes our unit of luminous sensation. … I propose to designate it a photon, using the English plural, photons. Symbolically, the letter ϕ will be assigned to it. … Each sensation is an accompaniment of a particular form of energy stimulus, i.e., of two, three or of four photons simultaneously discharged.

For example, a blue visual sensation was felt when $4\phi$ were discharged and the colours red and yellow corresponded to $2\phi$ and $2\phi + 3\phi$, respectively. Arguing that white light is caused by nine photons, he explained the complementary colours yellow and blue by the equality $(2\phi + 3\phi) + 4\phi = 9\phi$. Although he described his light quanta as parcels of radiant energy *hv*, he did not subscribe to or even mention Einstein's theory. At any rate, his photons were quite different from the quanta producing photoelectrons. Joly's photon made even less of an impact than Troland's. When the name was revived and caught on a few years later, no one recalled the name of the Irish physicist; or, for that matter, the name of the American psychologist.

**8. Conclusion**

While many words in physics can be traced far back in time, "photon" is a neologism coined only in the twentieth century. It is tempting to associate the name with Einstein's light quantum hypothesis, but the photon was originally introduced with a different meaning. The name was first suggested in 1916 and then independently five years later, in both cases as a unit related to the illumination of the eye and the resulting sensation of light. In 1925-1926 it reappeared in the French scientific literature, first in a physiological context and next in the context of physical optics.

Probably unaware of his predecessors, G. N. Lewis reinvented the name in the fall of 1926 for what he called "a kind of atom." Although this fifth photon was also quite different from Einstein's light quantum, this time the name (but not Lewis' concept) caught on. Coming at the right time it was considered an apt substitute for the older name.

**Acknowledgments**: I am indebted to Christophe Blondel for calling my attention to the early appearance of "photon" in French scientific literature (Section 4). Olivier Darrigol has kindly provided me with information concerning F. Wolfers.

**References**

Bacciagaluppi, Guido and Antony Valentini 2009. *Quantum Theory at the Crossroads: Reconsidering the 1927 Solvay Conference*. Cambridge: Cambridge University Press.

Band, William 1927. "Prof. Lewis' 'light corpuscles'." *Nature* **120**: 405-406.

Barten, Peter G. J. 1999. *Contrast Sensitivity of the Human Eye and its Effects on Image Quality*. Bellingham, Washington: International Society for Optical Engineering.

Berberan-Santos, Mário N. 2001. "Pioneering contributions of Jean and Francis Perrin to molecular luminescence." In *New Trends in Fluorescence Spectroscopy: Applications to Chemical and Life Sciences*, edited by Bernard Valeur and Jean-Claude Brochon. Berlin: Springer, pp. 7-33.

Birtwistle, George. 1928. *The New Quantum Mechanics*. Cambridge: Cambridge University Press.

Burchfield, Joe D. 1990. *Lord Kelvin and the Age of the Earth*. Chicago: University of Chicago Press.

Coffey, Patrick 2008. *Cathedrals of Science: The Personalities and Rivalries that Made Modern Chemistry*. Oxford: Oxford University Press.

Compton, Arthur H. 1928a. "Discordances entre l'expérience et la théorie électromagnétique du rayonnement." In *Électrons et Photons. Rapports et Discussions de Cinquième Conseil de Physique*, edited by Institut International de Physique Solvay. Paris: Gauthier-Villars, pp. 55-85.

Compton, Arthur H. 1928b. "Some experimental difficulties with the electromagnetic theory of radiation." *Journal of the Franklin Institute* **205**: 155-178.

Compton, Arthur H. 1928c. "X-rays as a branch of optics." *Journal of the Optical Society of America and Review of Scientific Instruments* **16**: 71-87.

Compton, Arthur H. 1929. "What things are made of." *Scientific American* **140** (February-March): 110-113, 234-236.

Comstock, Daniel F. and Leonard T. Troland 1917. *The Nature of Matter and Electricity: An Outline of Modern Views*. New York: Van Nostrand Co.

Dalton, John 1808. *A New System of Chemical Philosophy*. Manchester: R. Bickerstaff.

De Broglie, Louis 1925. "Recherche sur la théorie des quanta." *Annales de Physique* **3**: 22-138.

Dirac, Paul A. M. 1930. *The Principles of Quantum Mechanics*. Oxford: Oxford University Press.

Friedel, E. and F. Wolfers 1924. "Les variations de longueur d'onde des rayons X par diffusions et la loi de Bragg." *Comptes Rendus* **178**: 199-200.

Haas, Arthur E. 1928. *Materiewellen und Quantenmechanik*. Lepzig: Akademische Verlagsgesellschaft.

Henri, Victor and René Wurmser 1913. "Action des rayons ultraviolets sur l'eau oxygénée." *Comptes Rendus* **157**: 126-128.

Jammer, Max 1966. *The Conceptual Development of Quantum Mechanics*. New York: McGraw-Hill.

Joly, John 1921a. "A quantum theory of vision." *Philosophical Magazine* **41**: 289-304.

Joly, John 1921b. "A quantum theory of colour vision." *Proceedings of the Royal Society B* **92**: 219-232.

Kidd, Richard, James Ardini, and Anatol Anton 1989. "Evolution of the modern photon." *American Journal of Physics* **57**: 27-35.

Lewis, Edward S. 1998. *A Biography of Distinguished Scientist Gilbert Newton Lewis*. New York: Edwin Mellen Press.

Lewis, Gilbert N. 1926a. "The nature of light." *Proceedings of the National Academy of Science* **12**: 22-29.

Lewis, Gilbert N. 1926b. "The conservation of photons." *Nature* **118**: 874-875.

Mehra, Jagdish and Helmut Rechenberg 2000. *The Historical Development of Quantum Mechanics*, Vol. 6. New York: Springer.

Nudds, John R. 1986. "The life and work of John Joly (1857-1933)." *Irish Journal of Earth Sciences* **8**: 81-94.

Pais, Abraham 1982. *"Subtle is the Lord…": The Science and the Life of Albert Einstein*. Oxford: Oxford University Press.

Stranges, Anthony N. 1982. *Electrons and Valence: Development of the Theory, 1900-1925*. College Station: Texas A&M University Press.

Stuewer, Roger H. 1975a. *The Compton Effect: Turning Point in Physics.* New York: Science History Publications.

Stuewer, Roger H. 1975b. "G. N. Lewis on detailed balancing, the symmetry of time, and the nature of light." *Historical Studies in the Physical Sciences* **6**: 469-511.

Stuewer, Roger H. 2006. "Einstein's revolutionary light-quantum hypothesis." *Acta Physica Polonica B* **37**: 543-558.

Troland, Leonard T. 1916a. "The chemical origin and regulation of life." *The Monist* **24**: 92-133.

Troland, Leonard T. 1916b. "Apparent brightness; its conditions and properties." *Transactions of the Illuminating Engineering Society* **11**: 947-975.

Troland, Leonard T. 1917. "On the measurement of visual stimulation intensities." *Journal of Experimental Psychology* **2**: 1-33.

Troland, Leonard T. 1922. "The present status of visual science." *Bulletin of the National Research Council* **5**, Part 2: 1-120.

Walker, Charles T. and Glen A. Slack 1970. "Who named the –ON's?" *American Journal of Physics* **38**: 1380-1389.

Weyl, Hermann 1931. *The Theory of Groups and Quantum Mechanics*. New York: Dover Publications.

Wolfers, F. 1924. "Interférence par diffusion." *Comptes Rendus* **179**: 262-262.

Wolfers, F. 1925. "Sur un nouveau phénomène en optique; interférence par diffusion." *Journal de Physique* **6**: 354-368.

Wolfers, F. 1926. "Une action probable de la matière sur les quanta de radiation." *Comptes Rendus* **183**: 276-277.

Wolfers, F. 1929. *Transmutation des Éléments*. Paris: Éditions Scientifiques.

Wurmser, René 1925a. "La rendement énergétique de la photosynthèse chlorophylliene." *Annales de Physiologie et de Physicochimie Biologique* **1**: 47-63.

Wurmser, René 1925b. "Sur l'activité des diverses radiations dans la photosynthèse." *Comptes Rendus* **181**: 374-375.

Wurmser, René 1987. "Letter to the editor." *Photosynthesis Research* **13**: 91-93.